# nnSAM2: nnUNet-Enhanced One-Prompt SAM2 for Few-shot Multi-Modality Segmentation and Composition Analysis of Lumbar Paraspinal Muscles

Zhongyi Zhang *, Julie A. Hides, Enrico De Martino, Abdul Joseph Fofanah, Gervase Tuxworth

Zhongyi Zhang, MRes, School of Information and Communication Technology, Griffith University, Nathan, QLD 4111, Australia

Julie A. Hides, PhD, School of Allied Health, Sport and Social Work, Griffith University, Nathan, QLD 4111, Australia

Enrico De Martino, MD, PhD, Center for Neuroplasticity and Pain (CNAP), Department of Health Science and Technology, Aalborg University, Gistrup, 9260 North Jutland, Denmark

Abdul Joseph Fofanah, MSc, MEng, School of Information and Communication Technology, Griffith University, Nathan, QLD 4111, Australia

Gervase Tuxworth, PhD, School of Information and Communication Technology, Griffith University, Nathan, QLD 4111, Australia

* Corresponding author: Zhongyi Zhang

N44 2.34, Nathan Campus, Griffith University, Nathan QLD 4111, Australia

Phone: (61) 451857560

E-Mail: zhongyi.zhang@griffithuni.edu.au

**Manuscript Type:** Original Research

**Word Count for Text**: 2922

**Abbreviations:** AFL = Australian Football League, AGBRESA = Artificial Gravity Bed Rest—European Space Agency, DSC = Dice Similarity Coefficient, LBP = Low Back Pain, LES = Lumbar Erector Spinae, LPM = Lumbar Paraspinal Muscle, MF = Multifidus, nnSAM2 = no new SAM2, SAM2 = Segment Anything Model 2.

**Funding information**

This research received no specific grant from any funding agency in the public, commercial, or not-for-profit sectors.

**Data sharing statement**

Data analyzed during the study were provided by a third party. Requests for data should be directed to the provider indicated in the Acknowledgements.

# Key Points

- We present nnsam2 ("No-New SAM2," i.e., SAM2 without fine-tuning or modification), the first framework to integrate nnU-Net's 3D contextual awareness with SAM2's generalizability, achieving state-of-the-art few-shot multi-modality segmentation of lumbar paraspinal muscles.

- With only one annotated slice per dataset, nnsam2 achieved statistical comparability to expert measurements of muscle volume (multi-sequence MRI, multi-protocol CT), attenuation (multi-protocol CT), and fat ratio (Dixon MRI).

- Source code and public manual annotations are openly released to ensure transparency and reproducibility.

**Summary Statement**

With only a single annotated 2D slice per dataset, our method achieves state-of-the-art few-shot segmentation of lumbar paraspinal muscles across multi-modality imaging, yielding measurements of muscle volume (multi-sequence MRI, multi-protocol CT), attenuation (CT), and fat ratio (Dixon MRI) that are statistically comparable to expert references.

# Abstract

**Purpose**

To develop and validate No-New SAM2 (nnsam2) for few-shot segmentation of lumbar paraspinal muscles using only a single annotated slice per dataset, and to assess its statistical comparability with expert measurements across multi-sequence MRI and multi-protocol CT.

**Methods**

We retrospectively analyzed 1,219 scans (19,439 slices) from 762 participants across six datasets. Six slices (one per dataset) served as labeled examples, while the remaining 19,433 slices were used for testing. In this minimal-supervision setting, nnsam2 used single-slice SAM2 prompts to generate pseudo-labels, which were pooled across datasets and refined through three sequential, independent nnU-Net models. Segmentation performance was evaluated using the Dice similarity coefficient (DSC), and automated measurements—including muscle volume, fat ratio, and CT attenuation—were assessed with two one-sided tests (TOST) and intraclass correlation coefficients (ICC).

**Results**

nnsam2 outperformed vanilla SAM2, its medical variants, TotalSegmentator, and the leading few-shot method, achieving DSCs of 0.94–0.96 on MR images and 0.92–0.93 on CT. Automated and expert measurements were statistically equivalent for muscle volume (MRI/CT), CT attenuation, and Dixon fat ratio (TOST, $P < 0.05$), with consistently high ICCs (0.86–1.00).

**Conclusion**

We developed nnsam2, a state-of-the-art few-shot framework for multi-modality LPM segmentation, producing muscle volume (MRI/CT), attenuation (CT), and fat ratio (Dixon MRI) measurements that were statistically comparable to expert references. Validated across multimodal, multicenter, and multinational cohorts, and released with open code and data, nnsam2 demonstrated high annotation efficiency, robust generalizability, and reproducibility.

# 1. Introduction

Degeneration of the lumbar paraspinal muscles (LPM)—as seen on imaging as volume loss and fat infiltration—provides key quantitative markers for assessing musculoskeletal health. Such degeneration has been linked to chronic low back pain (LBP), a major global health concern that affected more than 619 million people worldwide in 2020 (1,2). In addition, LPM changes have been associated with spinal pathologies (e.g., spinal stenosis, spondylolisthesis, spondyloarthropathies), sarcopenia, metabolic and neuromuscular disorders, and prolonged immobilization, with imaging serving as a key tool for monitoring neuromuscular degeneration and assessing rehabilitation efficacy (3–7). Hence, imaging-based analysis is essential for advancing LPM research and clinical practice.

Magnetic resonance imaging (MRI) and computed tomography (CT) are common imaging modalities for LPM analysis (8). MRI provides superior soft-tissue contrast, facilitating detailed evaluation of muscle morphology and fat composition. Specifically, T1-weighted (T1W) and T2-weighted (T2W) MRI sequences yield high-resolution structural information, while Dixon sequences provide direct fat ratio estimation from fat–water separation (9,10). CT also supports anatomical assessment and fat ratio quantification, particularly when MRI is unavailable (8,11). However, current LPM analysis remains constrained by labor-intensive manual annotation, highlighting the need for automated segmentation.

Supervised deep learning can automate segmentation, but it requires training on large datasets with task-specific manual annotations, which are costly and time-consuming to generate (12–15). Few-shot medical image segmentation (FSMIS) alleviates this burden by learning from only a few labeled examples, yet most approaches remain tied to source-domain training and are generally limited to a single imaging modality. Cross-domain FSMIS (CD-FSMIS) extends this concept by adapting models from one modality (e.g., MRI) to another (e.g., CT), but it still relies on pretraining with a well-annotated source dataset before adaptation to a sparsely labeled target (16).

Recent foundation models have addressed this limitation by enabling general-purpose segmentation without task-specific training (17). Segment Anything Model 2 (SAM2) is among the most prominent and widely used in medical image segmentation (18). Unlike CD-FSMIS, which depends on source-domain pretraining constrained by dataset size and domain selection, SAM2 leverages large-scale, diverse pretraining to generalize to new domains without additional training. However, SAM2 lacks 3D contextual awareness in medical imaging. Medical fine-tuned SAM2 variants (19,20) attempt to mitigate this limitation, they reintroduce the need for task-specific annotations and are vulnerable to overfitting.

In this study, we present No-New SAM2 (nnsam2), a segmentation framework that combines the generalization capacity of vanilla SAM2 with the 3D contextual awareness of nnU-Net—leveraging both strengths without modifying or fine-tuning the foundation model—and thereby addresses the limitations of CD-FSMIS and SAM2 fine-tuning. Using only a single annotated slice per dataset, nnsam2 achieved state-of-the-art multi-modality LPM segmentation. Validated across multicenter MRI and CT datasets, it showed statistical equivalence to expert assessments of muscle volume, attenuation, and fat ratio. By releasing open code and data, nnsam2 promotes reproducibility and provides a data-efficient paradigm for expert-level LPM assessment across diverse imaging modalities.

# 2. Methodology

**Study sample**

This retrospective study was conducted in accordance with the Declaration of Helsinki. A total of 1,219 scans from 762 participants were retrospectively collected across six datasets, including 943 MRI scans from 486 participants and 276 CT scans from 276 participants (Figure 1). All datasets were obtained from prior studies and used exclusively for secondary analysis in this study. The MRI datasets included two private sets—(i) Australian Football League (AFL) players collected from November 2014 to March 2015; (ii) participants from the 60-day Artificial Gravity Bed Rest study conducted by the European Space Agency (AGBRESA) collected from March 2019 to December 2019 (5)—and one public set of patients with symptomatic back (21). The CT datasets included two public sets: (i) 176 participants from TotalSegmentator (15) encompassing both healthy individuals and patients with diverse conditions (e.g., tumors, vascular disease, trauma, inflammation, bleeding), and (ii) 100 participants from the Whole Abdominal ORgan Dataset (WORD), , consisting of abdominal CT scans from cancer patients undergoing radiotherapy planning (22).

In this study, we used six datasets spanning multiple modalities (Figure 1). Among them, AFL and AGBRESA were private datasets with appropriate ethical approvals, while the other four were publicly available open-access datasets that did not require ethics approval. The AFL study was approved by the Human Research Ethics Committee of the host institution (GU reference number 2017/896). The AGBRESA study was approved by the Northern Rhine Medical Association's ethics committee (no. 2018143) and registered with the German Clinical Trials Register (DRKS00015677). Dataset and ethics details are provided in Supplementary Text 1.

**Figure 1.** Flow diagram of dataset selection and exclusions

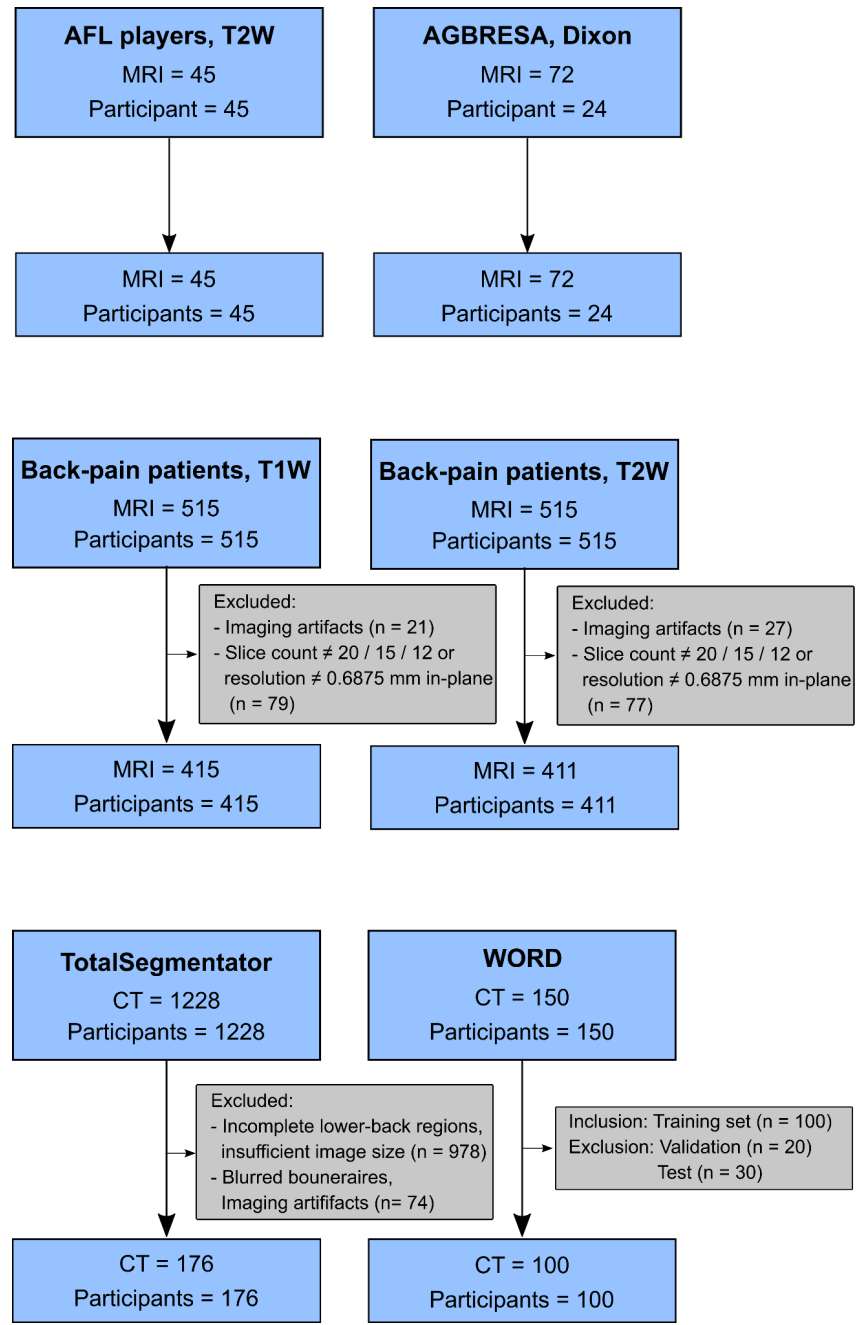

**Figure 1:** This diagram summarizes the exclusion process for MRI and CT datasets included in this retrospective study. For MRI datasets of back-pain patients, initial exclusions were due to imaging artifacts, followed by selection of axial series with 20, 15, or 12 slices and an in-plane resolution of 0.6875 × 0.6875 mm. For the TotalSegmentator dataset, automated screening retained 250 scans with complete L2–L5 segmentations, adequate in-plane size (≥256 × 256), and muscle labels >100 voxels. Manual review further excluded 74 low-quality scans (53 blurred margins, 20 artifacts, 1 vertebral encroachment), yielding 176 scans for final analysis. For the WORD dataset, only the training set was used, as its size was sufficient to represent the entire

dataset. T1W = T1-weighted; T2W = T2-weighted; AFL = Australian Football League; AGBRESA = Artificial Gravity Bed Rest Study—European Space Agency; WORD = Whole abdominal ORgan Dataset.

The public back-pain MRI dataset (21) was collected between September 2015 and July 2016 at Irbid Specialty Hospital in Jordan. For consistency, only scans with 20, 15, or 12 axial slices (0.6875 × 0.6875 mm in-plane resolution) were included, yielding 415 T1W and 411 T2W images (Figure 1). The TotalSegmentator dataset originally comprised 1,368 CT examinations collected at University Hospital Basel in 2012, 2016, and 2020 (15). Automated screening reduced this to 250 CTs by requiring valid L2–L5 segmentations, axial dimensions ≥256 × 256, and muscle labels >100 voxels. Subsequent manual review excluded 74 scans due to unclear muscle boundaries (n=53), artifacts (n=20), or vertebral encroachment (n=1), resulting in 176 CTs for analysis. For the WORD dataset (22), 100 training scans randomly sampled in the original study were used for evaluation. Other datasets were included as originally provided (Figure 1).

**Manual annotations**

Manual annotations focused on the L4–L5 region of the lumbar spine (spanning L3/L4 to L5/S1 disc levels), given its close association with fatty infiltration and LBP (23,24). Manual segmentations were firstly created by a trained annotator (3 months of LPM anatomy training) and independently reviewed and corrected by a physiotherapy professor (39 years of experience). The MF and LES muscles were segmented jointly to harmonize MRI and CT analysis, following the CT annotation convention (8,11,15). The paraspinal muscles in the TotalSegmentator dataset were manually re-segmented in our study to enhance annotation quality. Manual segmentation was performed in 3D Slicer (version 5.0.2, www.slicer.org). Annotations of the back-pain MRI datasets (826 scans, 13,633 slices) and two CT datasets (276 scans, 3,832 slices) are publicly available at https://github.com/johnnydfci/nnSAM2, with representative examples shown in Figure 2.

**Figure 2. Examples of manual and automated segmentations**

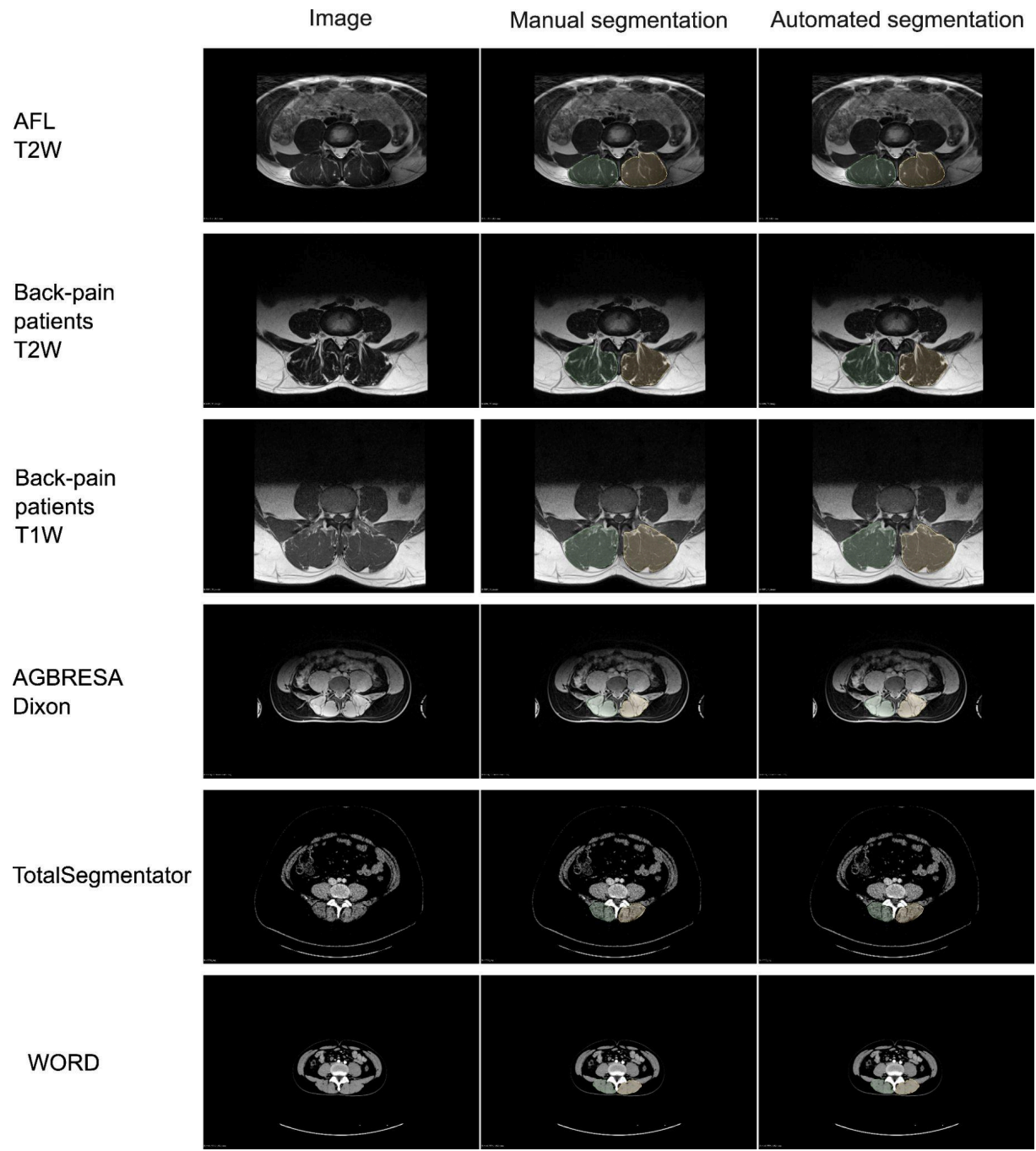

**Figure 2.** Examples of manual and automated segmentations across MRI and CT. Segmentations are displayed in green for the right LPM and yellow for the left LPM. LPM = lumbar paraspinal muscle; T1W = T1-weighted; T2W = T2-weighted; AFL = Australian Football League; AGBRESA = Artificial Gravity Bed Rest Study—European Space Agency; WORD = Whole Abdominal ORgan Dataset.

Following segmentation, we derived three quantitative metrics: muscle volume from MRI and CT, attenuation from CT, and fat ratio from Dixon MRI. Muscle volume was calculated as the product of the segmented voxel count and voxel spacing across MRI and CT. Fat infiltration ratio was assessed only in the AGBRESA dataset using paired Dixon MRI, defined as fat / (fat + water). Fat ratio calculation details are in Supplementary Text 2. Mean CT attenuation (HU) was computed as the average voxel intensity within each muscle mask.

For preprocessing, MRI pixel values were clipped to the 0.5th–99.5th percentiles (25). To harmonize through-plane resolution with MRI (~5 mm slice thickness), CT scans were downsampled along the head–foot axis (3× for TotalSegmentator, 2× for WORD). CT pixel values were clipped to [−30, 150] HU (26). Finally, all MR and CT images were rescaled to [0, 255] and converted to JPEG format (18), and then uniformly resized to 256 × 256 pixels.

**Deep learning method**

SAM2 (18) is a promptable framework that generates segmentation masks from user-defined prompts (points, boxes, masks). Pretrained on millions of images with a transformer-based encoder–decoder, it segments unseen images from prompts alone and outputs an IoU score indicating prediction confidence. In contrast, nnU-Net (27) is a fully supervised framework for biomedical segmentation, distinguished by its automated configuration pipeline and robust 3D contextual awareness. In this study, we present nnsam2 (Figure 3), a hybrid framework that integrates the generalization capacity of SAM2 with the 3D contextual awareness of nnU-Net. The source code is publicly available at https://github.com/johnnydfci/nnSAM2.

**Figure 3. Workflow of the nnsam2 framework**

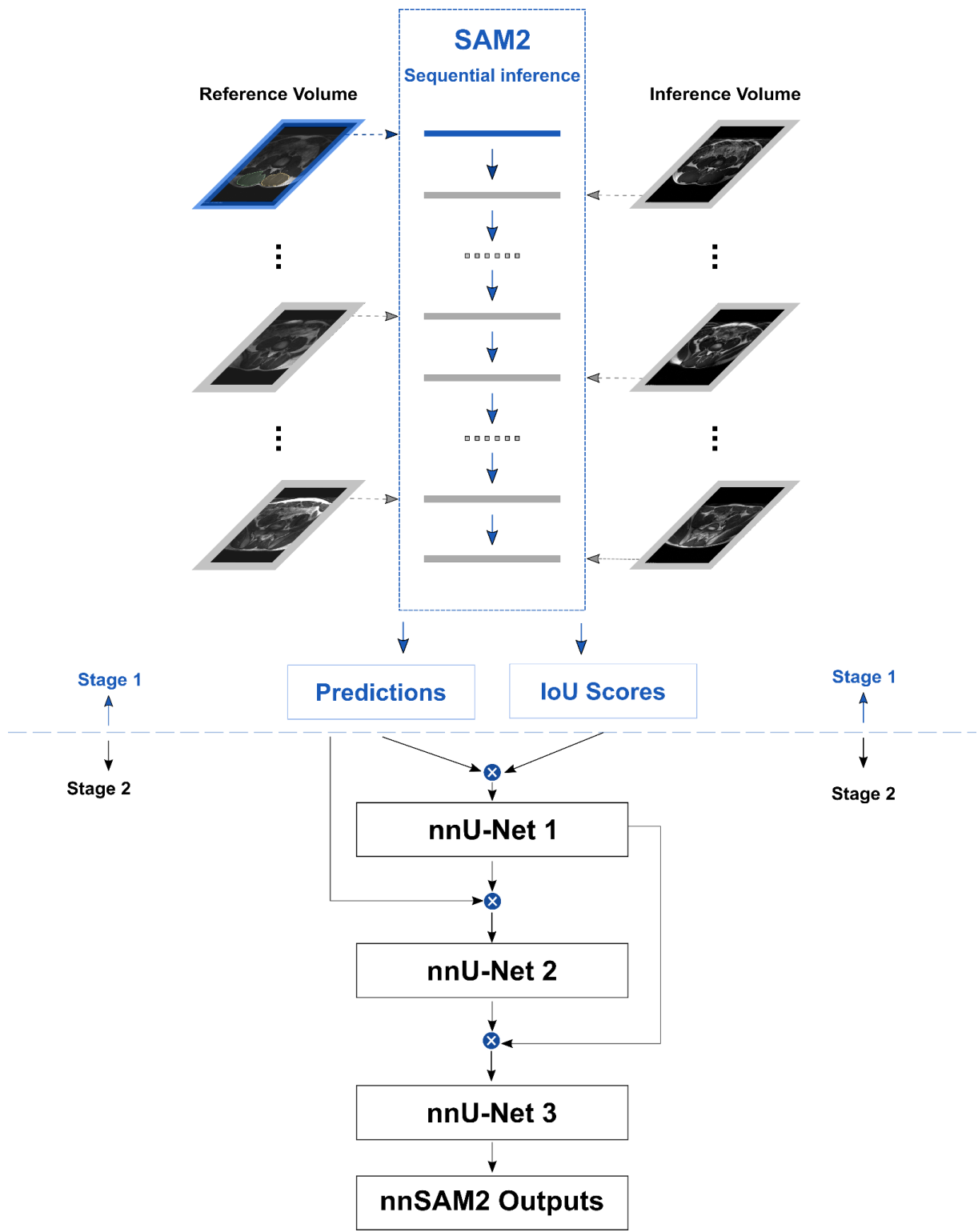

**Figure 3.** Workflow of the nnsam2 framework. In Stage 1, a single annotated slice (shown in blue) was used to prompt SAM2. The annotation was then propagated in an interleaved order between the reference and inference volumes, covering the L3/L4 to L5/S1 disc levels (for clarity, only the top, middle, and bottom slices are shown, while in practice 20+ slices may be included). SAM2 generated IoU scores as confidence estimates for all predictions, enabling the predictions to be used as pseudo-labels. In Stage 2, these pseudo-labels were refined through three sequential nnU-Net models with confidence-based filtering and anatomical smoothness constraints. In Step 1, the top 10% of masks by IoU at the dataset level and the top 2% at the slice level were pooled to train nnU-Net 1. In Step 2, predictions from nnU-Net 1 were retained if DSC with SAM2 predictions >0.90 and the cross-sectional area was ≤1.5× that of the superior slice, with the top 10% used to train nnU-Net 2. In Step 3, predictions from nnU-Net 2 were validated against nnU-Net 1 predictions with DSC >0.90 and a stricter smoothness threshold (≤1.25× superior slice), and the top 20% were used to train nnU-Net 3. Predictions from nnU-Net 3 served as the final nnsam2 outputs. The "X" symbols denote selection applied at each step, not mathematical multiplication. SAM2 = Segment Anything Model 2; nnSAM2 = no-new SAM2; nnU-Net = no-new U-Net; IoU = intersection over union; DSC = Dice similarity coefficient.

**nnsam2 Stage 1: SAM2-based Pseudo-labeling with IoU Confidence Scores.** In the first stage, pseudo-labels and IoU confidence scores were generated by SAM2 from a single annotated slice per dataset. SAM2 was designed for sequential inputs, propagating segmentation masks frame to frame. Although slices within a volume follow a natural top-to-bottom order, datasets may contain hundreds of volumes without any predefined inter-volume sequence, making manual definition of frame-to-frame propagation necessary. This was addressed by pairing a fixed reference volume with each inference volume and enforcing an interleaved slice order (Figure 3). The sequence alternated between reference and inference volume slices from top (L3/L4) to bottom (L5/S1). The reference volume was selected as the one with the smallest average feature distance across the L3/L4, L4/L5, and L5/S1 disc slices using DINOv2 features (28). From this reference volume, only the top slice was manually annotated and used as the prompting seed. In total, one reference slice per dataset was annotated, yielding only six annotated slices across the six test sets.

**nnsam2 Stage 2: Iterative refinement with nnU-Net.** Pseudo-labels generated by SAM2 were progressively refined using three sequential and independent nnU-Net models (Figure 3). At each stage, only high-confidence subsets of SAM2 predictions were retained—validated and filtered based on automated confidence scores and anatomical plausibility constraints—to ensure both reliability and smoothness across slices. The retained masks were successively used to train each subsequent nnU-Net model, with increasingly stringent selection criteria applied at every iteration to enhance label reliability. The outputs of the final nnU-Net served as the nnsam2 segmentations. Detailed refinement steps are provided in Figure 3 and Supplementary Text 3.

For benchmarking, we compared nnsam2 against (i) vanilla SAM2 (18) and its medical fine-tuned variants (19,20), representing promptable foundation-model approaches; (ii) FAMNet (16), the state-of-the-art CD-FSMIS method; and (iv) TotalSegmentator (12,15), a fully supervised segmentation baseline. All few-shot baselines used one annotated slice per dataset, while TotalSegmentator was applied externally without fine-tuning. Implementation details are provided in Supplementary Text 3.

**Statistical analysis**

The accuracy of artificial intelligence (AI)–based segmentations was evaluated using the Dice similarity coefficient (DSC) with corresponding standard deviations (SD). We additionally reported the minimal equivalence margins required for the two one-sided tests (TOST) of equivalence ($P < 0.05$), as well as mean absolute error (MAE) between AI-derived and manual measurements. The intraclass correlation coefficient (ICC) with 95% confidence intervals (CI) was also calculated using Pingouin (v0.5.5, Python package), based on a two-way random-effects model with absolute agreement for single measures. Agreement was further assessed using Bland–Altman analysis, with limits of agreement defined as the mean difference ± 1.96 SD.

All datasets were analyzed at the image level, except AGBRESA with three longitudinal time points per participant. For this dataset, we applied linear mixed-effects models (LMMs) to account for within-subject correlation following a prior study (29). TOST, ICC statistics were derived from the fitted LMMs to appropriately account for repeated measures. Detailed TOST, ICC, and model-based Bland–Altman analysis are provided in the Supplementary Text 4.

# 3. Results

**Participant characteristics**

A total of 1,219 scans from 762 participants were retrospectively collected (Table 1). The MRI datasets included 45 T2W scans from 45 AFL athletes, 72 Dixon scans from 24 participants in the AGBRESA study, and 826 scans from back-pain patients (415 T1W and 411 T2W). The CT datasets included 176 multi-protocol CTs from the TotalSegmentator dataset and 100 contrast-enhanced CTs from the WORD dataset.

**Table 1. Overview of MRI and CT datasets included in this study**

|  | Participants (no.) | Images (no.) | Slices (no.) | Age | Sex | Imaging protocol | Country | Source |
|---|---|---|---|---|---|---|---|---|
| AFL | 45 | 45 | 600 | 22.5 ± 2.8 | 45 male 0 female | T2W MRI | Australia | Private |
| Back-pain T2W | 411 | 411 | 6784 | Not provided | Not provided | T2W MRI | Jordan | Public |
| Back-pain T1W | 415 | 415 | 6849 | Not provided | Not provided | T1W MRI | Jordan | Public |
| AGBRESA | 24 | 72 | 1374 | 34±8; 32±10; 34±11 | 16 male 8 female | Dixon MRI | Germany | Private |
| Total segmentator | 176 | 176 | 2614 | 64.59 ± 14.18 | 110 male 66 female | Multi-protocol CT | Switzerland | Public |
| WORD | 100 | 100 | 1218 | 47 (28–75) | 63 male 37 female | Enhanced CT | China | Public |

**Table 1.** Age and sex information are presented as reported in the original studies. For the AGBRESA dataset, age is shown in three groups with eight participants each. AGBRESA = Artificial Gravity Bed Rest—European Space Agency; AFL = Australian Football League; T1W = T1-weighted; T2W = T2-weighted; WORD = Whole abdominal ORgan Dataset.

## Segmentation performance

On MR images, nnsam2 consistently outperformed all benchmarks (Table 2). In AFL T2W scans, it achieved a mean DSC of 0.95, exceeding SAM2 by ~0.03, FAMNet by >0.25, and TotalSegmentator by ~0.09–0.15. In AGBRESA Dixon scans, nnsam2 reached 0.94, again exceeding SAM2 by ~0.02, FAMNet by 0.21–0.29, and TotalSegmentator by ~0.09–0.15. The back-pain datasets showed the largest gains: in T1W, nnsam2 achieved 0.96–0.96 versus 0.93–0.94 for SAM2, ~0.67–0.68 for FAMNet, and ~0.83–0.85 for TotalSegmentator; in T2W, nnsam2 reached 0.96 compared with 0.93–0.94 for SAM2, ~0.73 for FAMNet, and ~0.82–0.84 for TotalSegmentator.

**Table 2. Automated segmentation performance on MRI and CT datasets**

|  | AFL T2W | | Back-pain T2W | | Back-pain T1W | | AGBRESA | | Total Segmentator | | WORD | |
| --- | --- | --- | --- | --- | --- | --- | --- | --- | --- | --- | --- | --- |
|  | (L) | (R) | (L) | (R) | (L) | (R) | (L) | (R) | (L) | (R) | (L) | (R) |
| SAM2 (18) | 0.92 ± 0.03 | 0.93 ± 0.02 | 0.94 ± 0.03 | 0.93 ± 0.02 | 0.94 ± 0.02 | 0.94 ± 0.01 | 0.92 ± 0.04 | 0.93 ± 0.03 | 0.91 ± 0.03 | 0.90 ± 0.08 | 0.90 ± 0.04 | 0.90 ± 0.03 |
| Med SAM2 (19) | 0.91 ± 0.03 | 0.90 ± 0.03 | 0.88 ± 0.05 | 0.88 ± 0.06 | 0.89 ± 0.05 | 0.90 ± 0.06 | 0.78 ± 0.26 | 0.80 ± 0.26 | 0.75 ± 0.26 | 0.67 ± 0.31 | 0.78 ± 0.16 | 0.81 ± 0.17 |
| Medical SAM2 (20) | 0.90 ± 0.03 | 0.90 ± 0.03 | 0.85 ± 0.07 | 0.87 ± 0.04 | 0.77 ± 0.12 | 0.79 ± 0.08 | 0.87 ± 0.06 | 0.89 ± 0.05 | 0.88 ± 0.04 | 0.86 ± 0.06 | 0.89 ± 0.05 | 0.89 ± 0.05 |
| FAMNet (16) | 0.69 ± 0.06 | 0.64 ± 0.16 | 0.67 ± 0.08 | 0.68 ± 0.13 | 0.73 ± 0.07 | 0.74 ± 0.11 | 0.73 ± 0.05 | 0.65 ± 0.10 | 0.73 ± 0.07 | 0.76 ± 0.06 | 0.61 ± 0.13 | 0.60 ± 0.23 |
| TotalSegmentator (12,15) | 0.80 ± 0.03 | 0.79 ± 0.03 | 0.84 ± 0.02 | 0.83 ± 0.03 | 0.84 ± 0.02 | 0.82 ± 0.03 | 0.85 ± 0.02 | 0.82 ± 0.02 | 0.87 ± 0.02 | 0.88 ± 0.02 | 0.89 ± 0.02 | 0.88 ± 0.02 |
| **nnsam2** | **0.95 ± 0.02** | **0.95 ± 0.01** | **0.96 ± 0.01** | **0.96 ± 0.01** | **0.96 ± 0.01** | **0.96 ± 0.01** | **0.94 ± 0.01** | **0.94 ± 0.01** | **0.92 ± 0.02** | **0.93 ± 0.02** | **0.92 ± 0.02** | **0.92 ± 0.03** |

**Table 2.** DSC results are reported as mean ± standard deviation for left and right paraspinal muscles. nnsam2 is highlighted in bold to denote the best-performing method under the strict one-slice-per-dataset supervision budget. AFL = Australian Football League; AGBRESA = Artificial Gravity Bed Rest—European Space Agency; DSC = Dice similarity coefficient; (L) = left muscles; (R) = right muscles; T1W = T1-weighted; T2W = T2-weighted; TotalSeg = TotalSegmentator; WORD = Whole Abdominal ORgan Dataset.

On CT, nnsam2 also outperformed all benchmarks. In the TotalSegmentator dataset, it achieved DSCs of 0.92–0.93, improving over SAM2 (0.90–0.91) by ~0.02, FAMNet

(0.73–0.76) by ~0.17–0.20, and the TotalSegmentator baseline (0.87–0.88) by ~0.05. In the WORD dataset, nnsam2 achieved 0.92–0.92, exceeding SAM2 (0.90–0.90) by ~0.02, FAMNet (0.60–0.61) by >0.30, and the TotalSegmentator baseline (0.88–0.89) by ~0.03–0.04. Medical variants of SAM2 again showed unstable performance, with Dice scores as low as ~0.67–0.78. Representative failure cases and their analyses are provided in Figure S1.

**Muscle quantification**

**Table 3. Muscle quantification: AI vs. manual annotations on MRI datasets**

| | AFL T2W Volume (mL) | | Back-pain T2W Volume (mL) | | Back-pain T1W Volume (mL) | | AGBRESA Volume (mL) | | AGBRESA Fat ratio | |
|---|---|---|---|---|---|---|---|---|---|---|
| Muscles | (L) | (R) | (L) | (R) | (L) | (R) | (L) | (R) | (L) | (R) |
| Manual | 225.08 ± 32.60 | 227.89 ± 33.47 | 391.04 ± 116.95 | 384.60 ± 115.11 | 385.65 ± 116.29 | 378.28 ± 114.45 | 155.79 ± 29.96 | 157.57 ± 31.41 | 0.1657 ± 0.0681 | 0.1751 ± 0.0669 |
| AI | 240.15 ± 32.68 | 236.20 ± 30.76 | 392.82 ± 114.12 | 384.01 ± 111.97 | 387.95 ± 114.91 | 384.09 ± 113.29 | 156.97 ± 29.88 | 158.36 ± 31.08 | 0.1654 ± 0.0622 | 0.1802 ± 0.0645 |
| TOST Min. δ | 7.93% | 4.85% | 0.63% | 0.34% | 0.73% | 1.70% | 2.06% | 1.96% | 2.09% | 4.86% |
| MAE | 15.47 | 11.24 | 6.31 | 6.38 | 5.42 | 7.48 | 4.35 | 5.07 | 0.0089 | 0.0108 |
| ICC [95% CI] | 0.86 [0.09–0.96] | 0.92 [0.70–0.97] | 1.00 [1.00–1.00] | 1.00 [1.00–1.00] | 1.00 [1.00–1.00] | 1.00 [0.99–1.00] | 0.94 [0.92–0.95] | 0.93 [0.91–0.94] | 0.97 [0.96–0.97] | 0.96 [0.95–0.97] |

**Table 3.** Muscle quantification: AI vs. manual annotations on MRI datasets. Results are reported as mean ± SD. Metrics include TOST Min. δ (%), MAE, and ICC with 95% CI. The minimum equivalence margin (%) indicates the smallest allowable difference between methods for statistical equivalence (TOST, p ≤ .05). (L) = left muscles; (R) = right muscles; HU = Hounsfield Unit; SD = standard deviation; TOST = Two One-Sided Tests; MAE = mean absolute error; ICC = intraclass correlation coefficient; CI = confidence interval; TotalSeg = TotalSegmentator; WORD = Whole Abdominal ORgan Dataset.

Automated and manual measurements of muscle volume, fat ratio, and CT attenuation (HU) were compared (Tables 3 and 4). On T1W scans, mean differences of muscle volume were <8 mL (MAE 7.48 mL right, 5.42 mL left), with equivalence margins <2% and ICCs of 1.00 [95% CI: 0.99–1.00]. On T2W scans, MAEs were similarly low (6.38–6.31 mL), with equivalence margins <1% and ICCs of 1.00. In the AFL cohort, volumes were slightly overestimated by 11–15 mL (MAE 11.24 mL right, 15.47 mL left), but still within equivalence margins (≤8%) and with strong agreement (ICC = 0.86–0.92). In the AGBRESA bed-rest study, right and left volumes showed excellent agreement (MAE 5.07 and 4.35 mL; ICC = 0.93–0.94).

**Table 4. Muscle quantification: AI vs. manual annotations on CT datasets.**

|  | TotalSegmentator Volume (mL) | | WORD Volume (mL) | | TotalSegmentator HU | | WORD HU | |
|---|---|---|---|---|---|---|---|---|
| Muscles | (L) | (R) | (L) | (R) | (L) | (R) | (L) | (R) |
| Manual | 113.09 ± 24.17 | 115.22 ± 24.57 | 100.71 ± 26.98 | 100.43 ± 26.80 | 23.13 ± 14.57 | 20.76 ± 14.93 | 35.25 ± 11.35 | 33.82 ± 11.93 |
| AI | 122.21 ± 25.16 | 122.92 ± 25.58 | 113.28 ± 28.89 | 111.28 ± 27.78 | 20.43 ± 14.22 | 18.88 ± 14.43 | 30.95 ± 11.59 | 30.46 ± 12.13 |
| TOST Min. δ | 8.60% | 7.26% | 13.25% | 11.67% | 12.34% | 9.94% | 13.07% | 11.07% |
| MAE | 9.25 | 7.79 | 12.66 | 10.85 | 2.71 | 1.97 | 4.33 | 3.52 |
| ICC [95% CI] | 0.92 [0.06–0.98] | 0.93 [0.25–0.98] | 0.90 [0.00–0.98] | 0.91 [0.01–0.98] | 0.98 [0.23–1.00] | 0.99 [0.79–1.00] | 0.92 [0.01–0.98] | 0.95 [0.29–0.98] |

**Table 4.** Muscle quantification: AI vs. manual annotations on CT datasets. Results are reported as mean ± SD. Metrics include TOST Min. δ (%), MAE, and ICC with 95% CI. The minimum equivalence margin (%) indicates the smallest allowable difference between methods for statistical equivalence (TOST, p ≤ .05). (L) = left muscles; (R) = right muscles; HU = Hounsfield Unit; SD = standard deviation; TOST = Two One-Sided Tests; MAE = mean absolute error; ICC = intraclass correlation coefficient; CI = confidence interval; TotalSeg = TotalSegmentator; WORD = Whole Abdominal ORgan Dataset.

For muscle volume on CT, nnsam2 modestly overestimated reference values. On the TotalSegmentator dataset, overestimations were 7.8 mL (right) and 9.3 mL (left), corresponding to equivalence margins of 7–9% and ICCs of 0.92–0.93. On the WORD dataset, overestimation was larger (10.9 mL right, 12.7 mL left), with equivalence margins of 11–13%, but agreement remained high (ICC = 0.90–0.91). Bland–Altman analysis further confirmed high agreement for volume measurements (Figure 4).

**Figure 4. Bland–Altman analysis of MRI and CT muscle volume measurements**

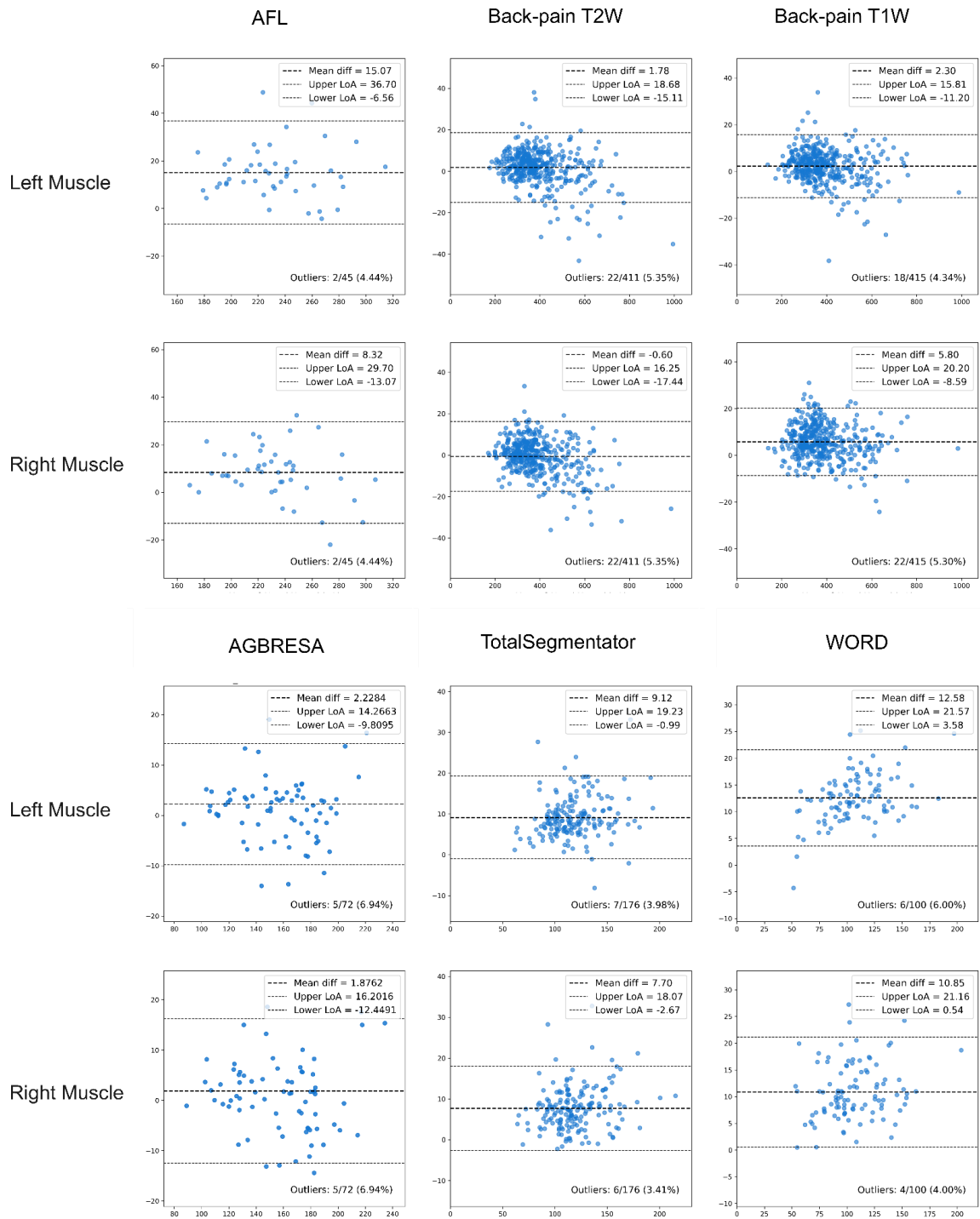

**Figure 4.** Bland–Altman plots of AI- and manual-derived muscle volume measurements from MRI and CT. The x-axis shows the mean of AI- and manual-derived values, and the y-axis represents their difference. The solid line indicates the mean difference, while dashed lines denote the 95% LoA. Data points outside the 95% LoA are considered outliers. AI = artificial intelligence; LoA = limits of agreement; MAE = mean absolute error; AFL = Australian Football League; AGBRESA = Artificial Gravity Bed Rest study by the European Space Agency; T1W = T1-weighted; T2W = T2-weighted; WORD = Whole Abdominal ORgan Dataset.

For the fat ratio on MRI, nnsam2 maintained statistical equivalence in the AGBRESA dataset. MAEs were 0.0108 (right) and 0.0089 (left), with equivalence margins ≤5% and ICCs of 0.96–0.97, confirming reliable quantification of muscle composition. For attenuation on CT, mean HU differences were 1.97 (right) and 2.71 (left), with ICCs of 0.98–0.99 in the TotalSegmentator dataset. In the WORD dataset, HU differences were slightly larger (3.5–4.3), but agreement remained strong (ICC = 0.92–0.95). Bland–Altman analysis further confirmed high agreement for fat ratio and HU measurements (Figure 5).

**Figure 5. Bland–Altman analysis of MRI fat ratio and CT attenuation measurements**

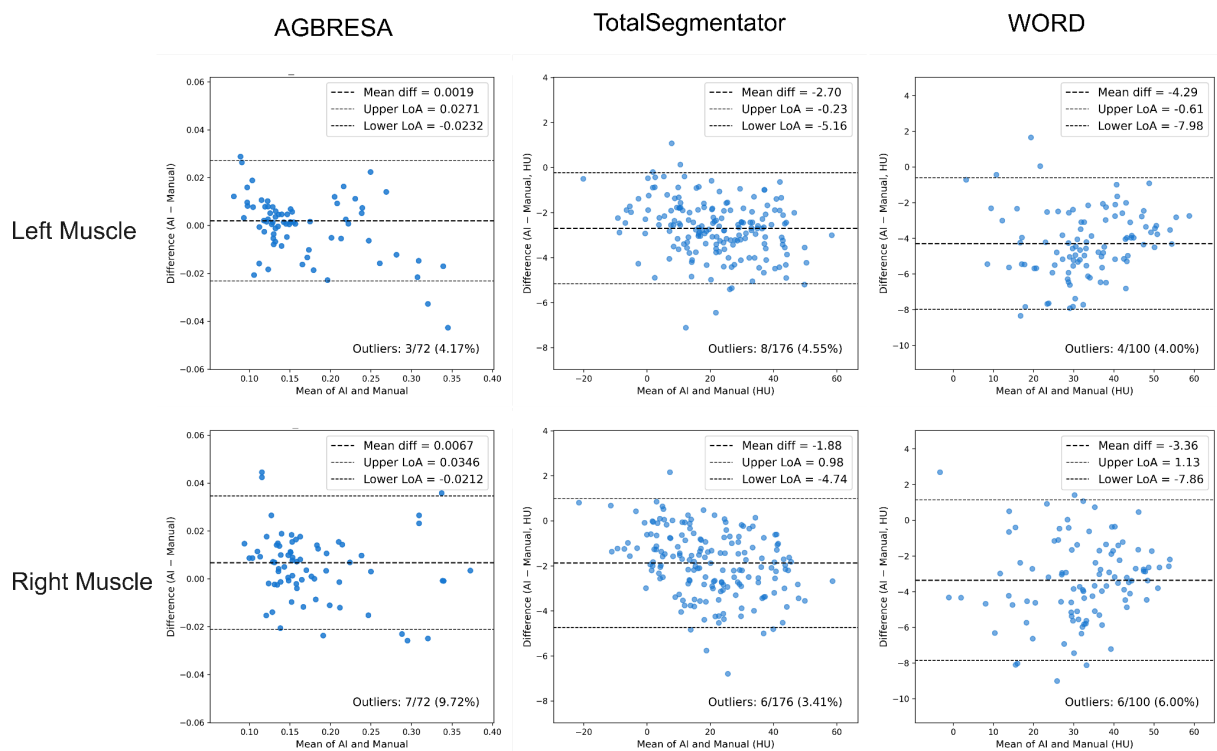

**Figure 5.** Bland–Altman analysis of MRI fat ratio and CT attenuation measurements. The x-axis shows the mean of AI- and manual-derived values, and the y-axis represents their difference. The solid line indicates the mean difference, while dashed lines denote the 95% LoA. Data points outside the 95% LoA are considered outliers. AI = artificial intelligence; HU = Hounsfield Unit; LoA = limits of agreement; MAE = mean absolute error; AGBRESA = Artificial Gravity Bed Rest study by the European Space Agency; WORD = Whole Abdominal ORgan Dataset.

## 4. Discussion

In this study, we present No-New SAM2 (nnsam2), a framework that achieves state-of-the-art performance in few-shot multi-modality LPM segmentation. The strength of nnsam2 does not stem from a new network architecture or from fine-tuning SAM2, but from systematically integrating SAM2's broad generalizability with nnU-Net's 3D contextual awareness. Beyond superior segmentation accuracy, nnsam2 achieved statistical equivalence to expert measurements with just six annotated slices across six diverse datasets, underscoring its value for clinical translation with minimal annotation. By releasing code and manual annotations (826 MRI scans with 13,633 slices and 276 CT scans with 3,832 slices), this research promotes reproducibility and accelerates future adoption.

We focused on the L4–L5 region of the lumbar spine (L3/L4 to L5/S1 disc level) because of its strong clinical relevance to LBP, degenerative spondylolisthesis, and prolonged immobilization (23,24,30,31). Muscles at this level are particularly susceptible to early fatty infiltration of the paraspinal muscles, linking muscle degeneration to disc pathology and spinal instability (32,33). Automated segmentation accuracy was lower on CT than on MRI, reflecting CT's limited ability to resolve LPM soft-tissue contrast rather than a deficiency of the algorithm. Accordingly, CT required wider equivalence margins of TOST due to lower contrast and higher inter-observer variability, whereas MRI allowed narrower margins given its clearer muscle boundaries.

We did not compare nnsam2 with single-modality FSMIS methods, as these approaches assume training and testing within the same modality. Such assumptions fail to generalize to cross-modal scenarios, making direct comparison neither fair nor clinically meaningful for our multi-modality segmentation task. Instead, nnsam2 was developed for cross-domain adaptation, aligning more closely with cross-domain FSMIS (CD-FSMIS) methods that aim to reduce annotation burden across modalities. Within this context, nnsam2 achieved markedly better performance than FAMNet, with more than 0.25 DSC improvement in several datasets, demonstrating stronger robustness to cross-domain variation. Moreover, nnsam2 advances beyond CD-FSMIS by leveraging SAM2 as a frozen foundation backbone, eliminating the need for an annotated source domain. This paradigm shift enables efficient LPM segmentation with minimal supervision, highlighting a new direction for the foundation model–based FSMIS.

Despite promising results, several limitations remain. First, manual annotations were generated by a trained expert and validated by a clinical professor; involving multiple experts in future work could further strengthen reliability. Second, we analyzed the LES and MF as a single class. This was necessary because LES and MF cannot be reliably distinguished on CT, and we applied the same grouping on MRI for consistency. Finally, nnsam2 relies on nnU-Net refinement, which may limit performance on very small datasets but provides clear advantages in large and heterogeneous cohorts. When data are insufficient, this limitation can be addressed by directly using SAM2 without nnU-Net enhancement.

In conclusion, we introduced nnsam2, a framework that achieves state-of-the-art few-shot, multi-modality LPM segmentation with only a single annotated slice per dataset. Beyond high segmentation accuracy, nnsam2 achieved statistical equivalence to expert measurements of muscle volume, CT attenuation, and fat ratio. By releasing source code and manual annotations, nnsam2 also promotes reproducibility and provides a data-efficient paradigm for foundation model–based LPM assessment.

# Acknowledgments

This research did not receive specific funding from public, commercial, or not-for-profit sectors. The study involved secondary analyses of data from prior research projects, including three public and two private datasets. We gratefully acknowledge the availability of the following public datasets:

Back-pain MRI — Mendeley Data: https://data.mendeley.com/datasets/k57fr854j2/2

TotalSegmentator (CT) — Zenodo: https://zenodo.org/records/10047292

WORD (CT) — GitHub: https://github.com/HiLab-git/WORD

The private datasets were obtained from earlier studies with funding sources unrelated to the present work. Although this study itself was unfunded, all datasets were used with appropriate ethical approval and informed consent. The AGBRESA study was supported by the German Aerospace Center, the European Space Agency (contract number 4000113871/15/NL/PG), and the National Aeronautics and Space Administration (contract number 80JSC018P0078), and was conducted at the *:envihab* research facility of the DLR Institute of Aerospace Medicine.

# References

1. Wesselink EO, Pool-Goudzwaard A, De Leener B, et al. Investigating the associations between lumbar paraspinal muscle health and age, BMI, sex, physical activity, and back pain using an automated computer-vision model: A UK Biobank study. Spine J. 2024; doi: 10.1016/j.spinee.2024.02.013.

2. Driscoll T, Jacklyn G, Orchard J, et al. The global burden of occupationally related low back pain: estimates from the Global Burden of Disease 2010 study. Ann Rheum Dis. 2014;73(6):975–981.

3. Feng N, Li W, Yu X, et al. Application of MRI imaging parameters in lumbar spine diseases: a review of the literature. Clin Radiol. Elsevier BV; 2025;80(106702):106702.

4. Hodges PW, Bailey JF, Fortin M, Battié MC. Paraspinal muscle imaging measurements for common spinal disorders: review and consensus-based recommendations from the ISSLS degenerative spinal phenotypes group. Eur Spine J. 2021;30(12):3428–3441.

5. Clément G, Rittweger J, Nitsche A, et al. Assessing the effects of artificial gravity in an analog of long-duration spaceflight: The protocol and implementation of the AGBRESA bed rest study. Front Physiol. Frontiers Media SA; 2022;13:976926.

6. De Martino E, Hides J, Elliott JM, et al. Lumbar muscle atrophy and increased relative intramuscular lipid concentration are not mitigated by daily artificial gravity after 60-day head-down tilt bed rest. J Appl Physiol. journals.physiology.org; 2021;131(1):356–368.

7. Wesselink EO, Hides J, Elliott JM, et al. New insights into the impact of bed rest on lumbopelvic muscles: a computer-vision model approach to measure fat fraction changes. J Appl Physiol. American Physiological Society; 2025;138(1):157–168.

8. Khil EK, Choi J-A, Hwang E, Sidek S, Choi I. Paraspinal back muscles in asymptomatic volunteers: quantitative and qualitative analysis using computed tomography (CT) and magnetic resonance imaging (MRI). BMC Musculoskelet Disord. Springer Science and Business Media LLC; 2020;21(1):403.

9. Han G, Jiang Y, Zhang B, Gong C, Li W. Imaging evaluation of fat infiltration in paraspinal muscles on MRI: A systematic review with a focus on methodology. Orthop Surg. Wiley; 2021;13(4):1141–1148.

10. Haueise T, Schick F, Stefan N, Machann J. Comparison of the accuracy of commercial two-point and multi-echo Dixon MRI for quantification of fat in liver, paravertebral muscles, and vertebral bone marrow. Eur J Radiol. Elsevier BV; 2024;172(111359):111359.

11. Trueb P, Getzmann JM, Ried E, Deininger-Czermak E, Garcia Schueler HI, Guggenberger R. Comparison of muscle fat fraction measurements in the lower spine musculature with non-contrast-enhanced CT and different MR imaging sequences. Eur J Radiol. Elsevier BV; 2022;150(110260):110260.

12. Akinci D'Antonoli T, Berger LK, Indrakanti AK, et al. TotalSegmentator MRI: Robust sequence-independent segmentation of multiple anatomic structures in MRI. Radiology. Radiological Society of North America (RSNA); 2025;314(2):e241613.

13. Wesselink EO, Elliott JM, Coppieters MW, et al. Convolutional neural networks for the

automatic segmentation of lumbar paraspinal muscles in people with low back pain. Sci Rep. 2022;12(1):13485.

14. Zhang Z, Li G, Wang Z, et al. Deep-learning segmentation to select liver parenchyma for categorizing hepatic steatosis on multinational chest CT. Sci Rep. Springer Science and Business Media LLC; 2024;14(1):11987.

15. Wasserthal J, Breit H-C, Meyer MT, et al. TotalSegmentator: Robust segmentation of 104 anatomic structures in CT images. Radiol Artif Intell. ncbi.nlm.nih.gov; 2023;5(5):e230024.

16. Bo Y, Zhu Y, Li L, Zhang H. FAMNet: Frequency-aware Matching Network for cross-domain few-shot medical image segmentation. Proc Conf AAAI Artif Intell. Association for the Advancement of Artificial Intelligence (AAAI); 2025;39(2):1889–1897.

17. Bian Y, Li J, Ye C, Jia X, Yang Q. Artificial intelligence in medical imaging: From task-specific models to large-scale foundation models. Chin Med J (Engl). Ovid Technologies (Wolters Kluwer Health); 2025;138(6):651–663.

18. Ravi N, Gabeur V, Hu Y-T, et al. SAM 2: Segment Anything in Images and Videos. arXiv [cs.CV]. 2024. http://arxiv.org/abs/2408.00714.

19. Ma J, Yang Z, Kim S, et al. MedSAM2: Segment anything in 3D medical images and videos. arXiv [eess.IV]. 2025. http://arxiv.org/abs/2504.03600.

20. Zhu J, Hamdi A, Qi Y, Jin Y, Wu J. Medical SAM 2: Segment medical images as video via Segment Anything Model 2. arXiv [cs.CV]. 2024. http://arxiv.org/abs/2408.00874.

21. Sudirman S. Lumbar Spine MRI Dataset. Mendeley; 2019. doi: 10.17632/K57FR854J2.2.

22. Luo X, Liao W, Xiao J, et al. WORD: A large scale dataset, benchmark and clinical applicable study for abdominal organ segmentation from CT image. Med Image Anal. 2022;82:102642.

23. Shi L, Yan B, Jiao Y, et al. Correlation between the fatty infiltration of paraspinal muscles and disc degeneration and the underlying mechanism. BMC Musculoskelet Disord. Springer Science and Business Media LLC; 2022;23(1):509.

24. Gu H, Hong J, Wang Z, et al. Association of MRI findings with paraspinal muscles fat infiltration at lower lumbar levels in patients with chronic low back pain: a multicenter prospective study. BMC Musculoskelet Disord. Springer Science and Business Media LLC; 2024;25(1):549.

25. Ma J, He Y, Li F, Han L, You C, Wang B. Segment anything in medical images. Nat Commun. 2024;15(1):654.

26. Tsai K-J, Chang C-C, Lo L-C, Chiang JY, Chang C-S, Huang Y-J. Automatic segmentation of paravertebral muscles in abdominal CT scan by U-Net: The application of data augmentation technique to increase the Jaccard ratio of deep learning: The application of data augmentation technique to increase the Jaccard ratio of deep learning. Medicine (Baltimore). Ovid Technologies (Wolters Kluwer Health); 2021;100(44):e27649.

27. Isensee F, Jaeger PF, Kohl SAA, Petersen J, Maier-Hein KH. nnU-Net: a self-configuring method for deep learning-based biomedical image segmentation. Nat Methods.

nature.com; 2021;18(2):203–211.

28. Oquab M, Darcet T, Moutakanni T, et al. DINOv2: Learning robust visual features without supervision. arXiv [cs.CV]. 2023. http://arxiv.org/abs/2304.07193.

29. Zhang Z, Hides JA, De Martino E, Millner J, Tuxworth G. Multicenter validation of automated segmentation and composition analysis of lumbar paraspinal muscles using multisequence MRI. Radiol Artif Intell. Radiological Society of North America (RSNA); 2025;(e240833):e240833.

30. Köhli P, Schönnagel L, Hambrecht J, et al. The relationship between paraspinal muscle atrophy and degenerative lumbar spondylolisthesis at the L4/5 level. Spine J. Elsevier BV; 2024;24(8):1396–1406.

31. Da W, Jian Q, Evan J, et al. Quantitative analysis of relationship between paraspinal muscle degeneration and degree of Degenerative lumbar spondylolisthesis. Spine (Phila Pa 1976). Ovid Technologies (Wolters Kluwer Health); 2025;10.1097/BRS.0000000000005270.

32. Özcan-Ekşi EE, Ekşi MŞ, Turgut VU, Canbolat Ç, Pamir MN. Reciprocal relationship between multifidus and psoas at L4-L5 level in women with low back pain. Br J Neurosurg. Informa UK Limited; 2021;35(2):220–228.

33. Seyedhoseinpoor T, Taghipour M, Dadgoo M, et al. Alteration of lumbar muscle morphology and composition in relation to low back pain: a systematic review and meta-analysis. Spine J. Elsevier; 2022;22(4):660–676.

# Supplements

**Text 1: MRI and CT dataset description and imaging parameters**

**Dataset 1: AFL players**

This dataset originates from a prior observational project that examined both the cross-sectional dimensions of trunk musculature and sagittal spinal alignment in professional athletes competing in the Australian Football League (AFL). The investigation was conducted under ethics approval from the university's Human Research Ethics Committee (GU Ref No: 2017/896). All players provided written informed consent before participating. While the entire team was invited, individuals were excluded if they opted out of imaging or presented contraindications for MRI, including claustrophobia, metallic piercings, cochlear or dental implants, or other medical devices considered unsafe in the scanner.

Forty-five male footballers from a single AFL club took part at the beginning of the competitive season. The group included 18 midfielders, 18 defenders, and 9 ruckmen. Their mean age was 22.47 ± 2.79 years, mean height 188.63 ± 6.66 cm, and mean body mass 87.24 ± 8.5 kg. On average, the players had competed in the league for 4.27 ± 2.94 years. Information about recent low back pain (LBP) was collected by questionnaire: 28 reported having LBP during the previous week, while 17 indicated no recent symptoms.

Imaging was conducted with players positioned supine and relaxed. Lumbar MRI scans were acquired using a Siemens Magnetom Verio 3T system (Siemens, Erlangen, Germany). Each session lasted approximately 15 minutes and included axial T2-weighted sequences spanning multiple lumbar intervertebral levels. Acquisition settings were: repetition time 7610 ms, echo time 87 ms, flip angle 120°, field of view 380 mm, and one average per scan.

**Dataset 2: Back-pain patients**

This dataset is derived from a publicly released collection of 515 patients who presented with symptomatic LBP. Data were acquired between September 2015 and July 2016 at Irbid Specialty Hospital in Jordan. Each subject underwent lumbar MRI examinations with both T1-weighted (T1W) and T2-weighted (T2W) protocols. The resource is distributed under a Creative Commons Attribution 4.0 International License and is freely accessible at https://data.mendeley.com/datasets/k57fr854j2/2. For the present study, we standardized inclusion by selecting only cases with 20, 15, or 12 axial slices at 0.6875 × 0.6875 mm resolution, yielding 415 T1W and 411 T2W scans.

Axial series typically covered the lowest three lumbar intervertebral discs (IVDs), extending to the lumbosacral junction. To match the curvature of the spine, slices through the terminal disc were angled along the disc plane, whereas slices at other levels were obtained in parallel. Each IVD was usually represented by four to five images encompassing both superior and inferior aspects. While coverage most commonly extended from L2 to L5, certain patients had up to 20 slices when additional vertebral levels were included.

Scanning was generally performed with participants positioned head-first and supine, though a minority were imaged feet-first. Acquisition time ranged from 15 to 45 minutes. In some instances, repeat scans of the same individual were available, either acquired on separate occasions or within a short interval of days.

**Dataset 3: AGBRESA study**

This dataset was obtained from the Artificial Gravity Bed Rest (AGBRESA) study conducted by the European Space Agency, which investigated the effects of prolonged head-down tilt (HDT) bed rest on intramuscular lipid content (ILC) of lumbar muscles over 60 days, and evaluated whether artificial gravity (AG) could mitigate these changes. Ethical approval was granted by the Ethics Committee of the Northern Rhine Medical Association, Düsseldorf

(Ärztekammer Nordrhein, No. 2018143), and the trial was registered with the German Clinical Trials Register (DRKS00015677).

Twenty-four healthy volunteers (16 men and 8 women) were enrolled and randomly assigned to one of three groups: (1) control (CTRL, n = 8), (2) continuous artificial gravity (cAG, n = 8; 30 minutes of uninterrupted AG daily), and (3) intermittent artificial gravity (iAG, n = 8; six 5-minute AG bouts per day). Baseline demographics were comparable across groups: CTRL (6 men, 2 women; age 34 ± 8 years; height 177 ± 7 cm; weight 79 ± 13 kg), cAG (5 men, 3 women; age 32 ± 10 years; height 173 ± 8 cm; weight 72 ± 10 kg), and iAG (5 men, 3 women; age 34 ± 11 years; height 174 ± 11 cm; weight 71 ± 5 kg).

The protocol included a 14-day baseline data collection (BDC) period before the 60-day strict −6° HDT bed rest, followed by 13 days of monitored recovery and rehabilitation. MRI was performed at three time points: two days before bed rest (BDC), on day 59 of bed rest, and after 13 days of recovery. At each visit, two Dixon sequences (water and fat) were acquired, yielding six MRI scans per participant across the study.

All MRIs were obtained on a Siemens Magnetom Vision 3T scanner (Siemens, Erlangen, Germany). Participants were positioned supine with knees and hips slightly flexed and supported by a pillow. Imaging covered the region from T12 to the sacrum with a T1-weighted Dixon sequence, producing 64 transverse slices. Acquisition parameters included: slice thickness 4 mm with 20% interslice gap, TR 7.02 ms, TE1 2.46 ms, TE2 3.69 ms, flip angle 5°, field of view 400 × 400 mm², in-plane resolution 1.0 × 1.0 mm², and total scan time of approximately 5 minutes. Both in-phase and out-of-phase fat–water images were reconstructed into separate water-only and fat-only volumes.

**Dataset 4: Totalsegmentator**

The TotalSegmentator dataset was first described in its original publication as version 1, which comprised 1,204 CT examinations collected in 2012, 2016, and 2020 at the University Hospital Basel. These scans were annotated for 104 anatomical structures (27 organs, 59 bones, 10 muscles, and 8 vessels), enabling applications such as volumetry, disease characterization, and preoperative or radiotherapy planning. Since the cases were retrospectively drawn from routine clinical CT, the dataset captured broad variability across patient demographics, pathologies, scanners, acquisition protocols, and institutions. Ethics approval was waived by the Ethics Committee Northwest and Central Switzerland (EKNZ BASEC Req-2022–00495).

To maximize diversity, the original cohort was randomly sampled, and scans with incomplete coverage, uncorrectable artifacts, or severe abnormalities were excluded. All images were resampled to 1.5-mm isotropic resolution. The collection represented data from eight institutions and 16 scanners (predominantly Siemens), covering multiple contrast phases (non-contrast, arterial, portal venous, late, and dual-energy). Among included patients, 404 had no abnormalities, while 645 presented with conditions such as tumors, vascular disease, trauma, inflammation, or bleeding.

In this study, we used the openly released version 2, which expands the dataset to 1,228 CT scans annotated for 117 anatomical structures, thereby increasing coverage of clinically relevant classes. Like version 1, the scans originate from routine clinical practice and thus generalize well across different institutions, scanners, and disease contexts. Version 2 is freely available at https://zenodo.org/records/10047292.

**Dataset 5: WORD**

The WORD dataset consists of 150 contrast-enhanced abdominal CT scans collected from 150 patients undergoing radiotherapy planning at a single center in China. All images were acquired on a Siemens CT system, without the use of appearance enhancement techniques.

Each study comprises 159–330 axial slices (512 × 512 pixels), with an in-plane resolution of 0.976 × 0.976 mm² and a native slice thickness of 2.5–3.0 mm, providing high spatial detail.

For our analysis, CT volumes were resampled to twice the original voxel spacing in the through-plane direction, giving an effective slice thickness of approximately 5 mm in the coronal orientation to maintain consistency with the MRI datasets. The WORD dataset is openly accessible and can be downloaded from the authors' repository at https://github.com/HiLab-git/WORD.

**Text 2: Manual measurement of fatty infiltration ratio**

After manual segmentation, the fatty infiltration ratio was calculated on Dixon MRI images as fat divided by the sum of fat and water signals. This computation was performed voxel by voxel, so that each voxel's fat and water intensity values contributed directly to the ratio. Such a voxel-wise approach provides a fine-grained assessment of intramuscular fat distribution rather than relying on averaged regional estimates. Among all datasets, only AGBRESA included paired Dixon fat and water images, making it suitable for this analysis. Segmentation was performed only on the Dixon-water images, as each participant had a paired water–fat image set in which every voxel in the water image corresponded exactly to the same voxel in the fat image, allowing a single mask to be applied to both. Each fat–water pair yielded one ratio per participant, and in total, 144 Dixon images produced 72 paired measurements of fatty infiltration.

**Text 3: Deep learning method and benchmarking**

In the nnsam2 framework, pseudo-labels were generated by SAM2 in Stage 1 and progressively refined through three independent nnU-Net models in Stage 2 (Figure 3). At each step of Stage 2, confidence-based filtering and anatomical plausibility constraints were applied to improve label quality and maintain anatomical smoothness across slices. The detailed refinement steps are as follows:

**Refinement Step 1 (1st nnU-Net training).** Pseudo-labels generated by SAM2 were filtered at two levels. At the dataset level, the top 10% of masks ranked by IoU were retained, while at the slice level the top 2% were preserved to ensure coverage of anatomically difficult slices. The union of these subsets formed the training set for the 1st nnU-Net. Although pseudo-labels were generated and ranked within each dataset, the selected subsets from all datasets were pooled together to train a single model, thereby encouraging cross-dataset generalization.

**Refinement Step 2 (2nd nnU-Net training).** Predictions from the 1st nnU-Net were validated against the predictions of SAM2. A prediction was retained only if its mean DSC with SAM2 exceeded 0.90 and its cross-sectional area was no larger than 1.5 times that of the superior slice, enforcing anatomical smoothness. From the validated set, the top 10% of predictions were used to train the 2nd nnU-Net.

**Refinement Step 3 (3rd nnU-Net training).** Predictions from the 2nd nnU-Net were validated against those from the 1st nnU-Net. The same DSC > 0.90 threshold was applied, but a stricter smoothness criterion was enforced, requiring the cross-sectional area to be no greater than 1.25 times that of the superior slice. From this refined pool, the top 20% of predictions were selected to train the 3rd nnU-Net.

**Final Output.** The trained 3rd nnU-Net served as the final nnsam2 predictor, and its outputs were used in benchmarking and downstream statistical analyses.

**Rationale for threshold selection**

Thresholds for pseudo-label selection and validation were chosen empirically to balance label quality with sufficient training coverage:

**Top 10% at dataset level (Refinement Step 1):** Retains only the most reliable SAM2 predictions while ensuring adequate samples for robust training. Pilot experiments showed that 5% yielded too few masks (underfitting), whereas 20% introduced excessive noise.

**Top 2% at slice level (Refinement Step 1):** Accounts for slices with systematically lower IoU due to artifacts or ambiguous anatomy, ensuring such slices were still represented in training.

**DSC > 0.90 (Refinement Steps 2–3):** Values above 0.90 are generally considered excellent agreement in medical segmentation. Relaxing to 0.85 admitted noisy masks, while 0.95 excluded too many valid masks and reduced generalization.

**Area ≤ 1.5× superior slice (Refinement Step 2):** Allowed natural anatomical variation while filtering out outliers. A looser cutoff (2.0×) admitted oversized masks; a stricter cutoff (1.25×) was too restrictive at this stage.

**Area ≤ 1.25× superior slice (Refinement Step 3):** A stricter smoothness constraint applied in the final refinement to stabilize inter-slice continuity. This cutoff eliminated noisy enlargements while preserving true variability.

**Top 20% for final training (Refinement Step 3):** Provided sufficient diversity while maintaining high confidence. Smaller pools (10%) reduced robustness; larger pools (30%) reintroduced noise.

**nnU-Net configuration**

The nnU-Net (version 1.7.1) served as the segmentation backbone within the nnsam2 framework. It automatically adapted preprocessing, architecture, training, and postprocessing settings to each dataset, ensuring robust performance across modalities. Training was performed on a Linux workstation (Ubuntu 18.04) with an NVIDIA RTX 3060 GPU for 1,000 epochs, using an initial learning rate of 0.01. A hybrid Dice–cross-entropy loss was applied to balance class overlap and pixel-wise accuracy. Default nnU-Net augmentations were used, except mirroring, which was disabled to avoid anatomical mislabeling. Postprocessing employed connected-component analysis to retain only the largest connected region per muscle class.

**Benchmarking methods**

For benchmarking, we compared nnsam2 against several representative methods. All few-shot baselines (SAM2, Med-SAM2, and FAMNet) were constrained to a single annotated slice per dataset, ensuring a fair comparison with nnsam2. In contrast, TotalSegmentator was applied externally as an "off-the-shelf" supervised model.

**Vanilla SAM2 and medical fine-tuned variants.** SAM2 was applied in a single-slice prompting mode, where one manually annotated slice per dataset served as the input prompt. The same strategy was used for vanilla SAM2 and two medical fine-tuned variants: MedSAM2 (Bowang Lab; https://github.com/bowang-lab/MedSAM2) and Medical-SAM2 (Imprint Lab; https://github.com/ImprintLab/Medical-SAM2). All models were run with their released pre-trained weights and no additional fine-tuning, ensuring consistency with the frozen SAM2 backbone used in nnsam2. Inference was performed slice-by-slice, and predictions were assembled into 3D volumes.

**FAMNet.** As the state-of-the-art cross-domain few-shot medical image segmentation (CD-FSMIS) framework, FAMNet was implemented using its official codebase

([https://github.com/primebo1/FAMNet](https://github.com/primebo1/FAMNet)). We used the publicly released pre-trained models for MRI→CT and CT→MRI transfer, which were originally trained on one source modality and applied to another. Since our evaluation covered both MRI and CT domains, both transfer models were applied accordingly. Each dataset was supported with a single annotated slice, in line with the strict one-slice-per-dataset supervision budget applied to nnsam2.

**TotalSegmentator.** As a fully supervised multi-organ segmentation model, TotalSegmentator was applied externally to both CT and MRI datasets without any fine-tuning or retraining. Predictions were restricted to the lumbar paraspinal muscle classes. The CT version, pre-trained on a large-scale, clinically diverse corpus, provides an approximate upper bound for supervised CT segmentation under real-world variability. Likewise, the recently released MRI version was applied directly to benchmark supervised performance in MRI. The official repository is available at [https://github.com/wasserth/TotalSegmentator](https://github.com/wasserth/TotalSegmentator).

**Text 4: Statistical analysis**

All computations were performed in Python 3.10. The main statistical routines were implemented with *statsmodels* (v0.13.5), *scipy* (v1.10.1), and *pingouin* (v0.5.5). Dice similarity coefficients (DSCs) were obtained using the *medpy* library (v0.4.0). DSCs were calculated at the image level for all datasets; for the AGBRESA Dixon dataset, values were additionally averaged per participant to account for repeated time points.

**Agreement analysis using linear mixed-effects models**

For the AGBRESA Dixon dataset, where each participant had repeated measurements across baseline, bed rest, and recovery, within-subject correlations were explicitly modeled using linear mixed-effects models (LMMs). Fixed effects were included for measurement method (AI vs. manual) and acquisition phase (three time points), while subject ID was entered as a random intercept to capture repeated observations within individuals. Random slopes were not added, as only 24 participants completed all three phases, limiting estimation stability. For muscle volume, scan sequence was initially tested as an additional fixed factor, but likelihood ratio tests showed no improvement, so it was removed for parsimony. Fat ratio analyses, available only from the AGBRESA Dixon scans, used the same fixed and random structure without sequence as a covariate.

**Statistical Equivalence Testing**

To evaluate whether AI- and manually derived measurements could be regarded as statistically equivalent, we employed the two one-sided test (TOST) framework. Estimates of the fixed effect ($\beta$) and their standard errors from the LMMs were used to calculate test statistics. Equivalence was concluded when the 90% confidence interval (CI) for $\beta$ lay entirely within the predefined equivalence bounds. Between the two one-sided tests, the larger p-value was taken as the decisive measure, with equivalence accepted if $p \leq 0.05$.

In addition to these fixed margins, we derived a minimum equivalence margin (Min. δ), defined as the narrowest allowable deviation from the manual reference that still satisfies the 90% CI condition for equivalence. Expressed as a percentage of the manual measurement, Min. δ provides an empirical threshold determined directly by the data. Thus, the analysis reports both perspectives: expert-defined tolerances that reflect clinical judgment, and Min. δ as a purely data-driven benchmark. The combination offers complementary insight, with Min. δ serves as the more objective and reproducible standard.

**Measurement Reliability: Intraclass Correlation Coefficient (ICC)**

Measurement reliability was further assessed using ICC, calculated from the variance components of the fitted LMMs. The ICC quantifies the proportion of total variance attributable to between-subject differences and is calculated as: ICC = $\sigma^2_{between}$ / ($\sigma^2_{between}$ + $\sigma^2_{within}$). $\sigma^2_{between}$ represents the variance of the random intercept (i.e., variability across individuals), and $\sigma^2_{within}$ captures residual variability due to measurement error and within-subject fluctuations.

**Model-Based Bland–Altman Analysis**

To evaluate agreement between AI and manual measurements while accounting for repeated observations, we adopted a Bland–Altman framework implemented through a linear mixed-effects model (LMM). In this model, the measurement difference (AI − manual) was specified as the dependent outcome, and participant ID was included as a random intercept to capture consistent subject-level deviations. The overall bias was obtained from the fixed intercept, while the limits of agreement (LoA) were calculated as: LoA = mean bias ± 1.96 × overall standard deviation, with the overall standard deviation estimated as the square root of the sum of residual (within-subject) variance and random intercept (between-subject) variance.

This formulation generalizes the traditional Bland–Altman method by incorporating both within-subject noise and systematic between-subject effects, producing more conservative and robust limits. Conventional approaches that rely only on residual error implicitly assume uniform agreement across individuals, ignoring the possibility that certain participants may consistently deviate in one direction. By explicitly modeling these subject-level biases via the random intercept, the total variance combines individual- and population-level variability, yielding a more realistic characterization of inter-method agreement.

# Figures

**Figure S1:** Automated segmentation examples of failures

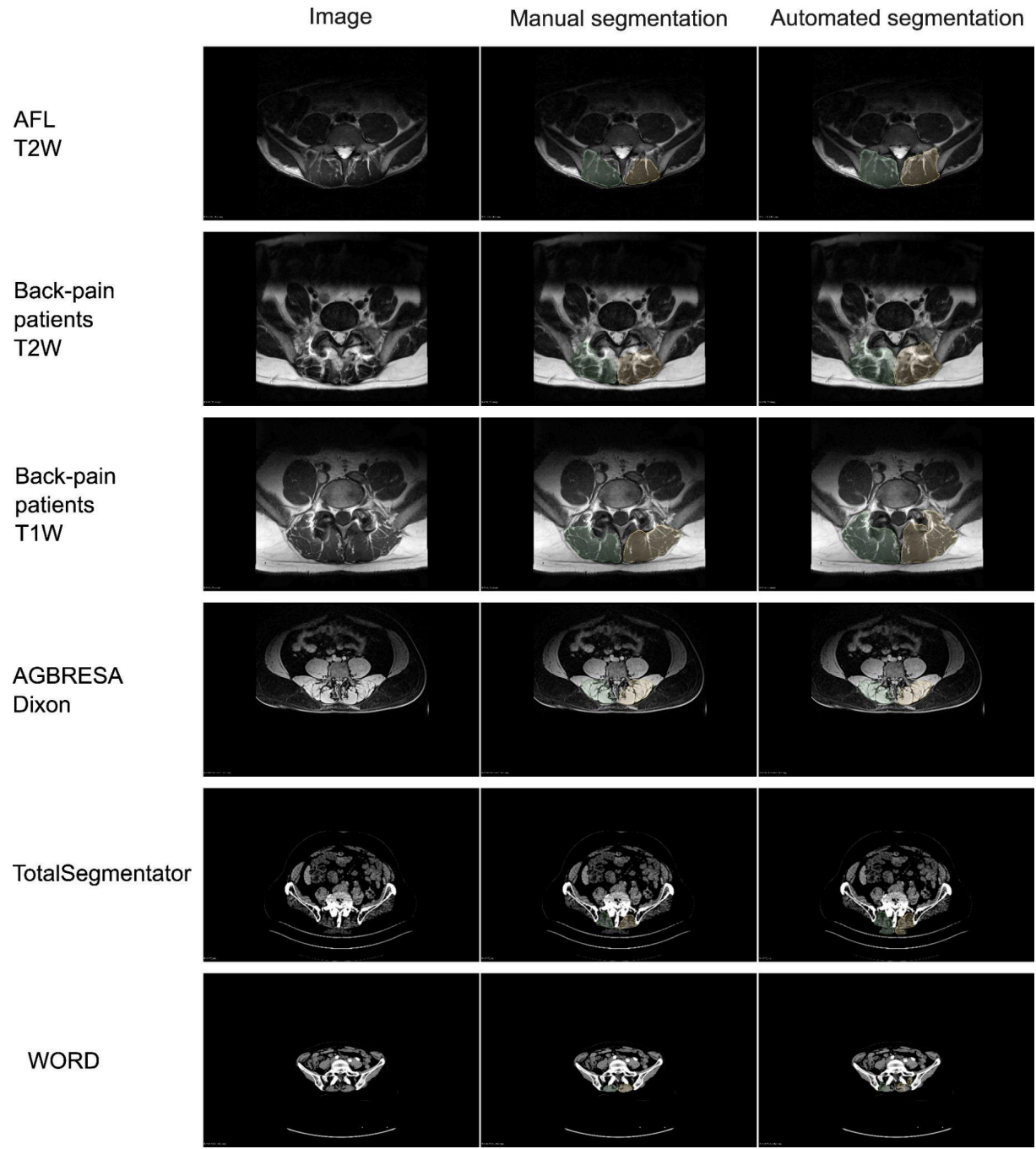

**Figure S1.** Examples of automated segmentation failures in MRI sequences (T1W, T2W, Dixon-water) and CT datasets. Segmentations are shown in green for the right LPM and yellow for the left LPM. Lower accuracy in these cases was mainly attributable to poor image resolution or indistinct muscle boundaries. LPM = lumbar paraspinal muscle; T1W = T1-weighted; T2W = T2-weighted; AFL = Australian Football League; AGBRESA = Artificial Gravity Bed Rest Study—European Space Agency; WORD = Whole Abdominal ORgan Dataset.